\documentclass[
    ,final            
  ]
  {aipproc}

\layoutstyle{6x9}

\begin{document}

\title{Disentangling age and  metallicity in distant unresolved
  stellar systems}


\author{M. Cantiello, E. Brocato and G.Raimondo}{
  address={INAF-OA Teramo, via Maggini, Teramo, Italy} }

\begin{abstract}
We present some results of an observational and theoretical study on
unresolved stellar systems based on the Surface Brightness
Fluctuations (SBF) technique. It is shown that SBF magnitudes are a
valuable tracer of stellar population properties, and a reliable
distance indicator. SBF magnitudes, SBF-colors, and SBF-gradients can
help to constrain within relatively narrow limits the metallicity and
age of the dominant stellar component in distant stellar systems,
especially if coupled with other spectro-photometric indicators.
\end{abstract}

\maketitle


\section{Introduction}
The detailed study of the properties of distant stellar systems relies
on the details to which the stellar population can be observed. The
observation of single stars is, in principle, the way to extract the
best information from a system. However, even with modern telescopes,
single stars can be observed only in nearby galaxies, and, typically,
at magnitudes significantly brighter than the main sequence. Here, we
discuss a technique that has proved to be very powerful to disentangle
the physical and chemical properties of distant, unresolved stellar
population, the Surface Brightness Fluctuations (SBF) method. Here, we
summarize the main characteristics of the technique - for a more
detailed description of the SBF method see G. Raimondo, and
J. P. Blakeslee's contributes to this Volume.

First introduced as a distance indicator for nearby elliptical
galaxies \citep{ts88}, the SBF method has been applied to Galactic
globular clusters at a few kpc, cluster ellipticals out to $\sim\,$150
Mpc, and numerous lenticulars, spiral bulges, and dwarf spheroidals at
intermediate distances \citep[e.g.][]{tonry01,mieske06a,biscardi08}.
By definition the SBF signal corresponds to the ratio of the 2$^{nd}$
to the 1$^{st}$ moment of the Luminosity Function (LF) of the stellar
population. Therefore, the dominant contribution to SBF comes from
bright stars which, acting as ``lighthouses'' on a smooth background
of faint stars, generate the great part of the fluctuation
signal. Further, the relative stellar luminosities depend on the
observed bandpass, so the $lighthouses$ at shorter wavelengths (hot
stars) are different from those at longer wavelength (cool stars), and
the main contributors to the SBF change according to the filter. This
makes SBF magnitudes and colors potential candidate to trace the
properties of stellar population in unresolved systems.

Based on such evidences we have carried out an extensive study of SBF,
and correlated photometric indicators, both on the theoretical and
observational point of view. The main results of these studies and
future perspectives are briefly outlined below.

\section{SBF: a different theoretical approach}
SBF magnitudes, as most other distance indicators, rely on the
calibration of the absolute magnitude. The most common SBF calibration
has been derived by \citet{tonry01}. Such empirical calibration uses
cluster membership to derive the slope, while the zeropoint relies on
6 galaxies with Cepheid's distances. Empirical calibrations, however,
are time consuming, and may suffer for zeropoint bias. For such
reasons, we started (since 1997) a campaign to derive theoretical SBF
calibrations in optical and near-IR bands, using the stellar
population synthesis method.

Taking advantage of the specific characteristic of the stellar
population synthesis code developed by the Teramo ``Stellar Population
Synthesis'' (SPoT) group \citep{brocato99,brocato00,raimondo02}, we
developed an original method to obtain SBF magnitudes from Simple
Stellar Populations, SSP, simulations
\citep{cantiello03,raimondo05}. The code uses the most reliable and up
to date physical inputs for stellar population synthesis. Some
ingredients of the models are: Initial Mass Function is from
\citet{scalo98} in the mass range $0.1 \leq M/M_{\odot} \leq
10$. Stellar evolution tracks from \citet{pietrinferni04}. The
horizontal branch morphology is fully reproduced taking into account
the effects due to age, metallicity, and the stellar mass spread due
to the stochasticity of the mass-loss phenomena along the RGB
\citep{brocato00}. The RGB mass-loss rate is evaluated according to
the \citet{reimers75} law. Thermal Pulses along the AGB phase are
simulated using the analytic formulations by \citet{wagenhuber98}. We
provide models computed assuming three different atmosphere models
(see \url{www.oa-teramo.inaf.it/spot} for more details, and models
download).

The accuracy of SPoT models has been proved against various observable
characteristics of resolved and unresolved stellar populations (CMD,
integrated colors, etc.), in all cases the agreement is
satisfactory. Then, we have checked SBF models versus available
optical and near-IR SBF data, obtaining either for local GC systems,
or for distant elliptical galaxies an excellent agreement with data
\citep{cantiello03,raimondo05}.  

\begin{figure}
  \includegraphics[height=.35\textheight]{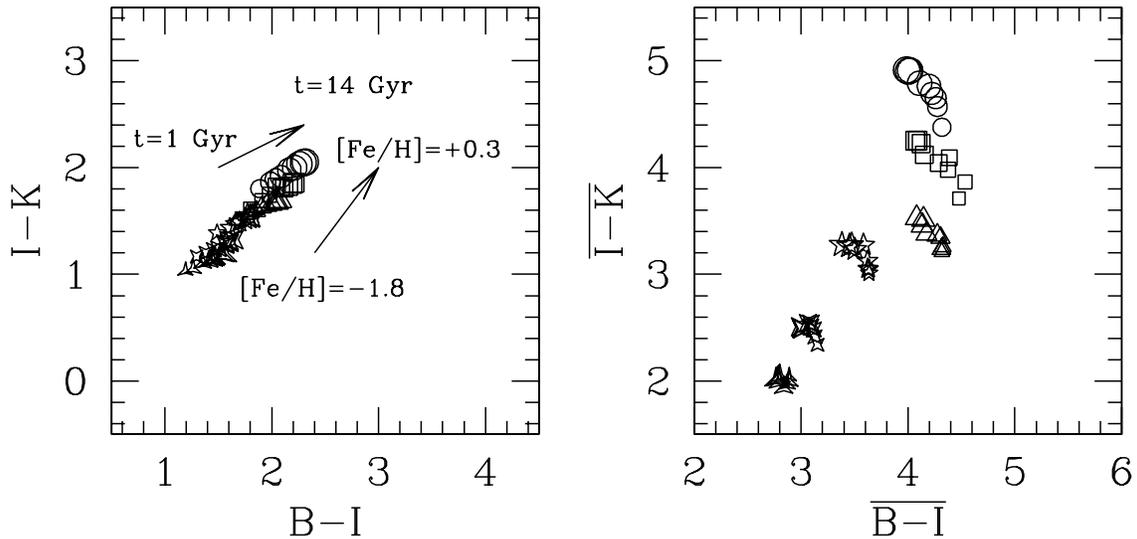} \caption{\small
$I{-}K$ versus $B{-}I$ integrated colors (left), and SBF-colors
(right), for SSP models with metallicity [Fe/H]
= $-$1.8, $-$1.3, $-$0.7, $-$0.3, 0.0, 0.3 dex (three, four and five
pointed stars, triangles, squares and circles, respectively), and age
$t=1$-14 Gyr (bigger squares refer to older ages). Models are from
\citep{raimondo05}. The arrows in the left panel indicate the
direction of larger [Fe/H] and older stellar populations, according to
labels.}
\label{colcol}
\end{figure}

The positive result of the tests on SPoT models, lead us to study more
in details how SBF magnitudes and colors can help to scrutinize the
properties of unresolved stellar populations. Figure \ref{colcol}
shows one particular application with SBF colors. Both panels show
color-color comparisons: integrated colors on the left, and
SBF-colors on the right. The figure shows how observational data
placed on the grid of models in the left panel would not significantly
constrain the mean age, $t$, or chemical composition, [Fe/H], of the
stellar population in the galaxy.  However, SBF-color data
\textit{can} constrain, within relatively narrow limits, the main
properties of the dominant stellar component. In other words the
``age-metallicity'' degeneracy is broken. This is also due to the fact
that the SBF measurement uncertainties can be below $\sim$ 0.1 mag,
less that the separation between models at different [Fe/H].

\section{SBF measurements}

Most of the observational work on SBF has been carried out based on the
needs of distance measurements, and little or no effort has been put
to secure good wavelength coverage on a single galaxy, so that a
homogeneous set of SBF measurements can be obtained. Using archival
ACS data, we have made an attempt to derive optical SBF colors for a
sample of $\sim$20 galaxies. The results of the study have been
encouraging on a twofold basis. First, the comparison of data to
models has shown a general good agreement on optical bands like V, and
I. Moreover, it has shown that for bluer bands, like B, the effect of
stellar population properties becomes non-negligible and, for example,
composite stellar population models have to be used instead of SSPs
\citep{cantiello07a}. Second, data to models comparisons have been
successfully used to constrain the properties of stars in
galaxies. Although optical SBF colors are not the first choice
colors to reliably constrain the $t$ and [Fe/H] of the stellar
system, the good matching of the stellar population properties derived
with SBF-colors and other derivations taken from literature has
demonstrated the feasibility of this kind of applications.

Furthermore, for what concerns SBF and stellar population properties,
we have recently demonstrated that radial SBF gradients can be
measured out to $\sim$30 Mpc, with present observing facilities, and
used as another important analysis tool \citep{cantiello05}. As an
example, using ACS data of 7 ellipticals, we found that the observed
color and SBF gradients are likely due to metallicity variations along
the galaxy radius, rather than to age variations. The additional step
of measuring SBF-color gradients can reveal subtle variations in
stellar population properties within the galaxy, so that the past
history of merging, interactions, or passive evolution can be
scrutinized from a promising new point of view.

\section{Future applications}
SBF and SBF-colors predictions from detailed stellar population
synthesis models show that specific applications of this method are
capable to lift the age-metallicity degeneracy which affects the study
of distant unresolved stellar systems. In particular, SPoT models
predict that the use of optical to near-IR SBF colors should be
preferred for stellar population studies.

In addition, present optical facilities have proved that SBF gradients
can be measured and used, for example, to understand how stellar
population properties change with radius in spheroidal galaxies. At
present no near-IR detection of SBF gradients exist. However, it is
well reasonable to expect that modern large telescopes equipped with
large format near-IR detectors can be used to detect near-IR SBF
gradients. Such possibility appears even more realistic with future
30-40 m class telescopes, or space-based facilities, like JWST or WSO
(the latter for the blue and UV wavelength regime). Such telescopes
will, in fact, allow to secure good wavelength coverage from UV to
near-IR (and possibly mid-IR) with high efficiency and spatial
resolution.

Coupling optical and near-IR SBF gradients, i.e. using SBF-color
gradients, will provide very accurate data to be used for the study of
how galaxies formed and evolved.


\begin{theacknowledgments}
Financial support for this work was partially provided by PRIN-INAF
2006 ``From Local to Cosmological Distances'' (P.I. Clementini G.)
\end{theacknowledgments}



\bibliographystyle{aipproc}   

\bibliography{cantiello}

\IfFileExists{\jobname.bbl}{}
 {\typeout{}
  \typeout{******************************************}
  \typeout{** Please run "bibtex \jobname" to obtain}
  \typeout{** the bibliography and then re-run LaTeX}
  \typeout{** twice to fix the references!}
  \typeout{******************************************}
  \typeout{}
 }

\end{document}